# Partial Discharges detection in 1 MV power supplies in MITICA experiment, the ITER Heating Neutral Beam Injector prototype

Marco Boldrin[a], Mattia Dan[a], Vanni Toigo[a,b], Loris Zanotto[a], Paolo Barbato[a,b], Lucio Baseggio[a], Manola Carraro[a,b], Raffaele Ghiraldelli[a,b], Enrico Zerbetto[a], Stefano Malgarotti[c], Alberto Rizzi[c], Giuseppe Rizzi[d], Hans Decamps[e], Hiroyuki Tobari[f]

[a]*Consorzio RFX (CNR, ENEA, INFN, Università di Padova, Acciaierie Venete SpA), Corso Stati Uniti 4, 35127 Padova, Italy*
[b]*Istituto per la Scienza e la Tecnologia dei Plasmi, CNR, Padova, Italy*
[c]*CESI S.p.A., Via Rubattino 54, I-20134 Milano, Italy*
[d]*CESI S.p.A. consultant*
[e]*ITER Organization, Route de Vinon-sur-Verdon, CS 90 046, 13067 St. Paul Lez Durance Cedex, France*
[f]*National Institute of Quantum and Radiological Science and Technology, Mukouyama 801-1, Naka 319-0193, Japan*

MITICA (Megavolt ITER Injector & Concept Advancement), the full scale prototype of ITER Heating Neutral Beam, is under realization at the Neutral Beam Test Facility (Padova, Italy). It is designed to deliver 16.5 MW to ITER plasma, obtained by accelerating negative Deuterium ions up to 1 MeV for a total ion current of 40 A and then neutralized.

MITICA Acceleration Power Supply is composed of several non-standard equipment, beyond industrial standard for insulation voltage level (-1 MVdc) and dimensions.

Voltage withstand tests (up to 1.265 MVdc) have been performed in five subsequent steps (from 2018 to 2019), according to the installation progress, after connecting equipment belonging to different procurements.

During integrated commissioning, started in 2021, two breakdowns occurred in a position of the HV plant not still identified, so they could be occurred either in air or in SF6. To identify the locations of possible weak insulation points, the existing diagnostics for partial discharge detection (the precursor of breakdowns) as a first step have been improved on air-insulated parts by consisting in a set of instrumentation, like capacitive probes and off-the-shelf instruments for AC application (acoustic and electromagnetic sensors).

The paper deals with the instruments qualification to assess their suitability for DC usage and then with the investigation performed in MITICA, in particular:

1) sensitivity assessment campaign, with artificially produced corona effect to identify the minimum threshold of each diagnostics

2) voltage application to MITICA plant, moving the instrumentation around equipment and increasing progressively the voltage looking for corona phenomena to identify possible weak insulation points.

## 1 Introduction

ITER plasma burning conditions and configuration control are achieved by additional heating systems including two heating and current-drive neutral beam injectors (HNBs). Each of them is designed to deliver 16.5 MW, obtained by accelerating negative Deuterium ions up to 1 MeV for a total ion current of 40 A (46 A of Hydrogen ions at 870 keV) and then neutralized, with a beam-on time lasting up to 1 hour [1]. To achieve such very demanding performances (so far never simultaneously attained) in time for ITER operation, a dedicated Neutral Beam Test Facility (NBTF) called PRIMA (Padua Research on Injector with Megavolt Acceleration) [2] is under realization in Padova (Italy), at Consorzio RFX premises. PRIMA hosts two experiments: SPIDER (Source for the Production of Ions of Deuterium Extracted from RF plasma), in operation since June 2018, is the full-scale test-bed of the negative Ion Source with 100 keV particle energy; MITICA (Megavolt ITER Injector & Concept Advancement) is the 1:1 scale prototype of the ITER HNB aimed at testing the production, extraction, acceleration and neutralization of H- and D- ions. MITICA Power Supply (PS) is a complex system, presenting a variety of insulation technologies (oil, $SF_6$, air, vacuum) and composed of several non-standard equipment beyond industrial standard for insulation voltage level (-1 MVdc) and dimensions (see Fig. 1):

- the Acceleration Grid Power Supply (AGPS, see detail 1 in Fig. 1) [3], feeding a voltage down to -1 MV dc across the electrodes of the electrostatic accelerator, composed of five -200 kV series connected DC Generators (DCG); each DCG consists of a three phase step up oil insulated transformer which high voltage output is rectified by a three phase diodes bridge contained inside a $SF_6$ pressurized tank; DCG output conductors are connected to a RC Filter unit (DCF, detail 2) via $SF_6$ gas insulated lines;

- a large -1 MV dc air insulated Faraday cage, the High Voltage Deck1 (HVD1, detail 3) [4][5], hosts ISEPS,

*author's email: marco.boldrin@igi.cnr.it*

- the Ion Source and Extraction Power Supply system [6] providing power to the Ion Source for production and extraction of the negative ions; ISEPS is fed by an Insulating Transformer (detail 4) [7] and its output conductors are routed from the HVD1 to the Transmission Line (TL) through an air-to-SF$_6$ Bushing (High Voltage Bushing Assembly, HVBA, detail 5); these equipment, insulated in air, are installed inside a High Voltage Hall (HVH) with controlled ambient conditions (air temperature, humidity and dust level) [5];
- finally, the 100 m long SF$_6$ (at 0.6 MPa abs) insulated TL (detail 6), composed of 3 subsequent sections (TL1 to TL3 [3]), connects AGPS and ISEPS to the beam source installed inside MITICA Vessel (detail 8) through the SF$_6$-to-vacuum HV Bushing (HVB, detail 7). Provisionally, during the insulation and integrated tests, the Short Circuiting Device (SCD) is installed in place of the beam source. The SCD, insulated in SF$_6$ at 0.13 MPa abs, is provided with five movable spark gaps, to simulate grids breakdown (BD); five fixed parallel spark gaps limit the voltage to the HVB admissible value (-1 MVdc in total, 200 kV for each stage).

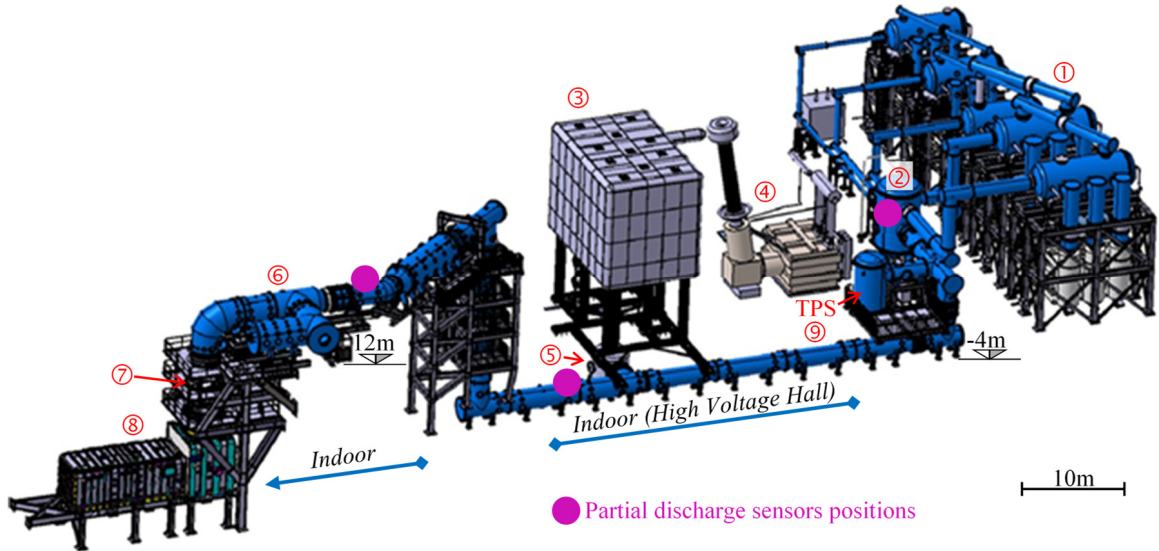

Fig. 1 Overall view of the MITICA PS system.

## 2  Insulation and integrated tests issues

The equipment, procured by European (EUDA) and Japanese (JADA) Domestic Agencies (DA's), performed successfully acceptance tests in Factories, either as a mockup of reduced dimensions (HVD1 [5]) or as a full scale sample ([3], HVBA [5]) before delivering at PRIMA site. MITICA Power Supply system has been installed from 2016 to 2019 and High voltage withstand tests (consisting of three consecutive voltage applications: 1.2 MV x 1 h, 1.06 MV x 5 h, after the 5 h application 5 peaks at 1.265 MVdc level [3]) have been performed in five subsequent steps from September 2018 to November 2019, according to the installation progress and after connecting equipment belonging to different procurements. Tests were carried out remotely, by applying voltage with the Testing Power Supply (TPS) generator (rated for 1.3 MV, 10 mA, detail 9 in Fig. 1) according to the following sequence:

1) test on DCGs (TL disconnected);
2) test on TL1+TL2+TL3 (from this test on DCGs disconnected);
3) test on HVD1+HVBA (only 1.2 MV x 1 h) for HVD1+HVBA partial acceptance; TPS connected through TL1, TL2+TL3 also connected;
4) test on HVD1+HVBA+Insulating Transformer; TPS connected through TL1, TL2+TL3 also connected;
5) test on HV Bushing, limited at 1 MVdc x 1 h to cope with its voltage withstanding capability; TPS connected through TL1, TL2+TL3+HVD1+HVBA+Insulating Transformer also connected, electrostatic termination (SCD) installed inside MITICA Vessel.

As far concerns the diagnostic, the SF$_6$ insulated equipment (mainly the TL) was provided with 3 embedded partial discharge (PD) detectors, based on ultrahigh frequency (UHF) principle and on a proprietary software for signal processing and evaluation, distributed along TL routing as shown in Fig. 1. The HVH was monitored by video surveillance system consisting of two fixed cameras in the visible range (namely "2+M" and "5" in Fig. 2, the former provided with ambient microphone) and one special ultraviolet (UV) [8] OFIL DayCor pan-and-tilt camera ("UV" in Fig. 2) to inspect the air insulated equipment and surroundings to identify possible sources of PD, the precursor of breakdowns. MITICA PS integrated commissioning started in 2021 by using the AGPS dummy load ("DL" in Fig. 2), consisting of a series of high voltage resistors in air with connections taps to be connected to the HVD1 for on-site testing of AGPS performance, temporarily installed inside the HVH just for this purpose. A pulse of 700 kV, lasting for 1000 s, was performed at a first attempt in such large NBI, an important result in view of ITER/DEMO; nevertheless, advancing with the test campaign two breakdowns occurred at higher voltage values somewhere in the HV plant, either in air or in SF$_6$, without clear evidence of the source provided by the existing set of diagnostics.

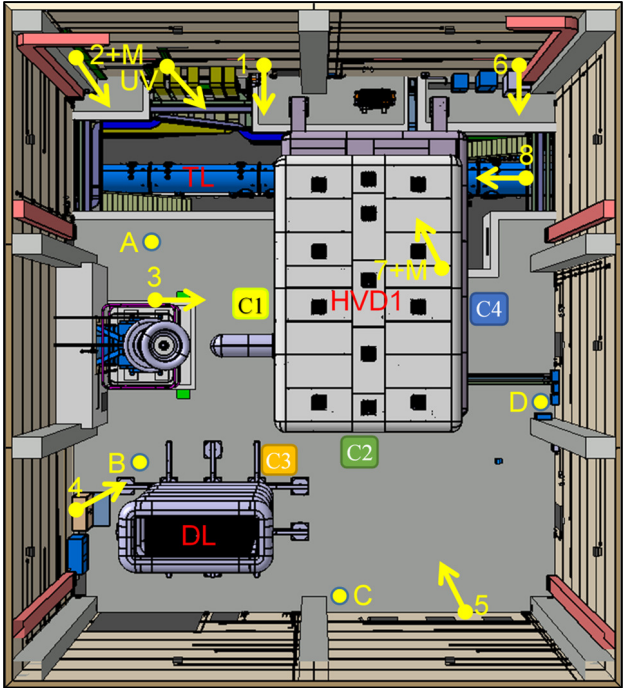

Fig. 2 Plan view of diagnostics layout in HVH.

## 3  Partial discharges detection improvement

### 3.1  Selection of the instrumentation

To identify the locations of air-insulated parts with weak insulation, a set of instruments suitable to detect corona under DC voltage has been selected. Corona effect is associated to light (in the visible and UV range) and sound emission, electric charge release (due to electric field variation) and ozone production. There are different off-the-shelf solutions for corona detection in AC applications, where corona pulses are cyclically produced and can be correlated to the voltage sine waveform (PD phase resolved pattern) to identify the type of defect generating the corona. For DC application, where corona/PD repetition rate is usually lower often with quite long time interval in between two consecutive pulses, the experience of the use of the same measuring systems is very limited and therefore check are necessary to verify instrumentation suitability. The following candidate instruments were identified:

a) Acoustic sensors: the chosen off-the-shelf solution was the NL Camera [9], equipped with a matrix of 124 microphones to catch sounds emitted by corona occurrence and instantly showing the sound source location on the camera's monitor; the instrument has been installed on a tripod, provided with a camera to display out of the HVH, segregated during the tests, on MITICA control room for online monitoring (see Fig. 3);

b) Electromagnetic sensor: the selected instrument (DOBLE PD measuring system type PDS100, provided with DA100 Directional Antenna [10]) searches for PD in the radio frequency (RF) range (50 MHz - 1000 MHz) and displays a "footprint" of the RF signal produced by the partial discharge source (corona are typically within 500 MHz); also in this case the instrument, with its antenna, has been supported by a tripod provided with camera to display the footprint;

c) capacitive field probe consisting of a voltage unit (i.e. the "low voltage" branch of a capacitive voltage divider which "high voltage" branch is the stray capacitance between the field probe and the electrode under voltage) sensing the electrode electric field variations originated either by changes in the applied voltage or because of the space charge in front of the probe; similar capacitive field probe has been used by CESI as a linear voltage divider [11], [12]. The probes for the tests have been built from a single square sheet of printed circuit board with copper coating on both sides, drilled in the centre to allow for connection of the two metallic sides to a RG-58 cable. Different samples were built, the final dimension of the probes chosen for the use were 53 cm x 61 cm, 9.5 nF of measured capacitance.

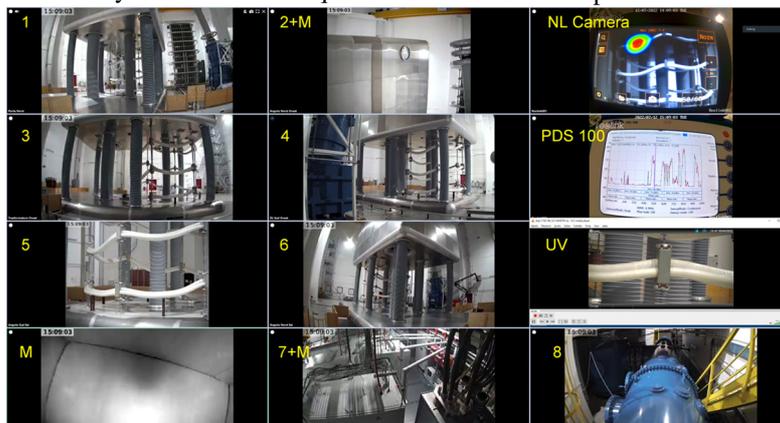

Fig. 3 Instruments and cameras online monitoring inside MITICA control room.

### 3.2  Instruments verification and field of view assessment

The first activity was aimed at verifying the correct operation of the instruments identified for corona detection in DC application. The NL camera (A), the PDS100 (B), the DayCor camera (C) and the capacitive field probe (D) were installed in front of a sharp point supplied by a negative DC voltage generator to produce corona effect (see Fig. 4).

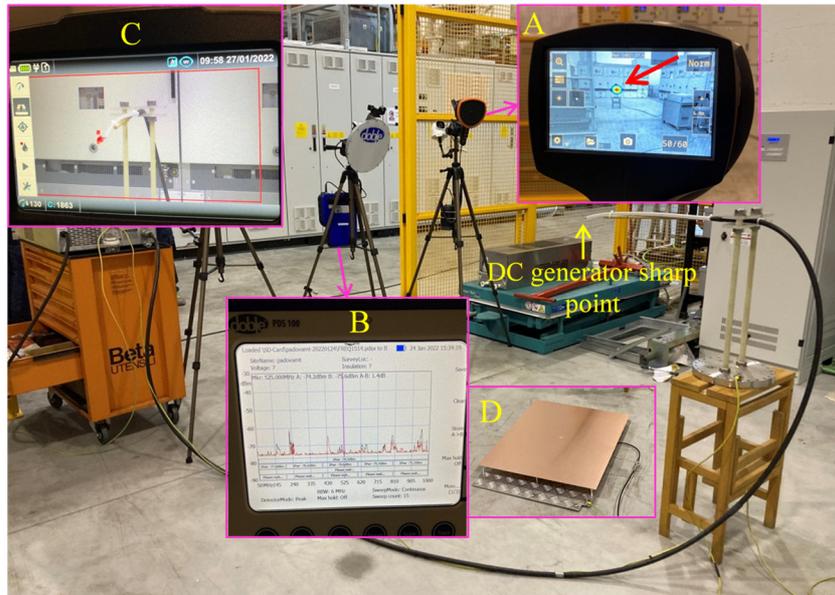

Fig. 4 Test setup for instruments verification. Displays show acquisitions at 5 m from the sharp point.

The instruments A, B and C, positioned at a distance of about 1.5 m from the PD source for a first scan, were later moved at about 5 m from the sharp point, supplied at -40 kV in both cases, and the measurements were repeated. All the instruments showed a similar sensitivity, detecting the "corona" inception at -27 kV whilst the extinction voltage was -20 kV for both the distances from the source. As visible in Fig. 4, the NL camera (A) identifies the corona source (sharp point) with a coloured area while the DayCor camera (C) indicates the source of defect with a red dots cloud superposed on the background ambient image. It has to be noticed that DayCor camera detects as well the natural activity present in air, therefore also in case of no corona occurrence some red dots could be present on the screen but they are widespread and their number is quite low. The PDS100 (B) screen shows instead the variation of the acquired signal curve (in black) with respect to the background noise (in red) recorded after instrument positioning before voltage application. The capacitive field probe (D) was firstly characterized by recording its output to a step electric field produced applying a DC voltage to a metallic sheet installed 50 mm above the capacitive probe. The DC generator step voltage was set at -250 V, Fig. 5 shows that the field probe signal decreases to zero voltage with the decay time (~ 10 ms) determined by the probe capacitance (9.5 nF) and the input resistance of the digital oscilloscope (1 MΩ). Once installed at about 1 m from the sharp point connected to the DC voltage generator (see Fig. 4) the capacitive field probe output signal remained (correctly) at zero voltage up to 20 kV, then increasing the applied voltage at corona occurrence the output signal raised to a negative value of few tens of millivolt. A more accurate analysis of the record had shown that this was due to the short time between two subsequent corona pulses, shorter than the characteristic discharge time of the capacitive probe, determining therefore a mean value slightly different from zero. The tests described showed the capability of the selected instruments to detect "corona" effect generated in DC field.

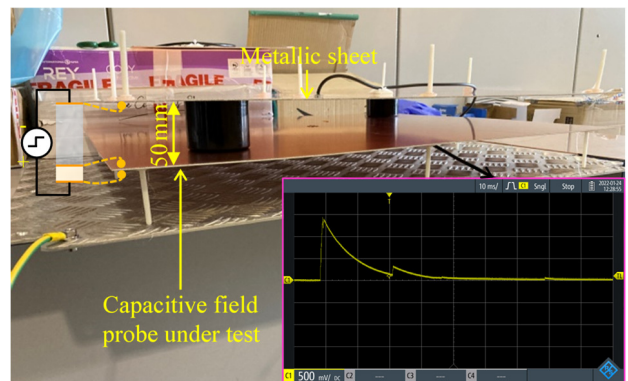

Fig. 5 Test setup for probe answer to a step electric field.

Additional tests were performed to determine NL camera and PDS100 field of view.

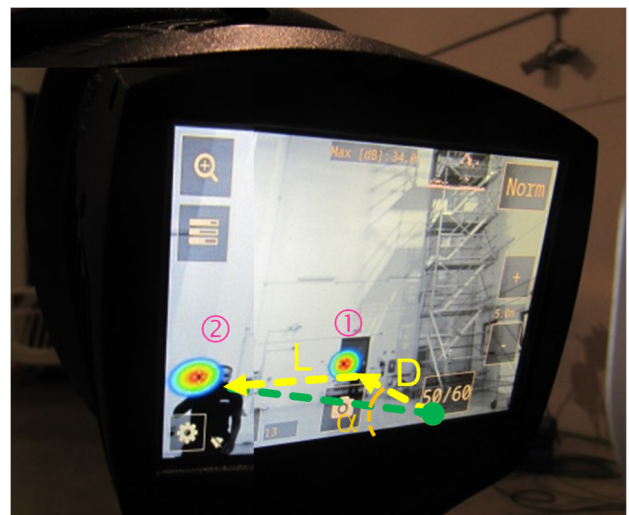

Fig. 6 Test to determine NL camera field of view.

This was done by using a commercially available piezoelectric gas lighter; once activated it generated several "small" sparks that produced sound and modified the electric field. For both instruments in turn, the lighter was aligned with the equipment axis at the maximum available distance "D" (about 14 m) in the HVH and hence the capability of the sensors to detect the event verified (see ① in Fig. 6 for NL camera). Then the lighter was moved horizontally out of the axis and the test repeated to detect the maximum horizontal distance "L" for which the sensors were able to detect the event (see ② in Fig. 6 for NL camera). The corresponding angle α was calculated; the field view of each sensor, equivalent to the double of α, was determined in 60° for NL camera and 130° for PDS100 respectively.

## 4 High voltage tests

### 4.1 Test procedure

The tests [13] were carried out energizing the plant with the TPS once disconnected the DCGs and the Insulating Transformer, therefore in a configuration quite similar to the voltage withstanding test 5 (see Section 2), limiting to -1 MV dc the maximum voltage application. Considering the huge space to monitor inside the HVH and the effective field of view of the instruments, the diagnostic equipment (NL camera and PDS 100) were installed to spot a limited area; in total 4 points were selected to monitor the complete installation (from "A" to "D" in Fig. 2). Fixed cameras in the visible field were increased to 6 (added cameras 1, 3, 4, 6 in Fig. 2) distributed along the HVH perimeter to cover the whole installation together with the DayCor camera. The following test procedure was defined as a guideline; depending on the test results, deviations are allowed upon agreement among Teams attending the tests:

1) with diagnostic equipment in position "A", to increase gradually the voltage with step of 100 kV from 0 up to 300 kV, voltage application to be maintained for at least 10 minutes for each step to detect corona activity; in absence of corona detection, to increase the voltage to the next step, otherwise reduce the voltage to zero and inspect the plant to identify the cause of corona or ionization and fix the issue before raise again the voltage;

2) the diagnostic instrumentation has then to be moved to the subsequent positions (from "B" to "D" in Fig. 2) and the voltage application sequence as per point 1 repeated; the same to be done also for the remaining positions;

3) once tested all the positions, the voltage has to be further raised, in step of 100 kV, from 300 kV to 600 kV and the tests repeated for all the positions;

4) after the completion of point 3, the voltage has to be raised from 600 kV to 700 kV and then, with step amplitude reduced to 50 kV, to 800 kV;

5) finally, the test sequence has to be completed raising gradually the voltage from 800 kV to 1000 kV with the instruments positioned subsequently in the four observation points inside the HVH.

### 4.2 Tests results

A first test session was carried out from January 24$^{th}$ until January 28$^{th}$ 2022. Only two capacitive field probes, C1 and C3 in Fig. 2, were used and their signals acquired through a digital oscilloscope remotely controlled. Few minutes after having reached 500 kV, after an irregular TPS current behaviour, the voltage dropped to zero. No indication of "corona" or other phenomena were noticed during the tests. Fig. 7 shows the last available data at 500 kV: it has to be evidenced that NL camera always pointed out one (two) coloured areas on the HVD1 bottom side (depending from the observation point) but it has been clarified that such noise was generated by HVD1 cooling plant and was not correlated with any PD generation, being present in the same location also without voltage application.

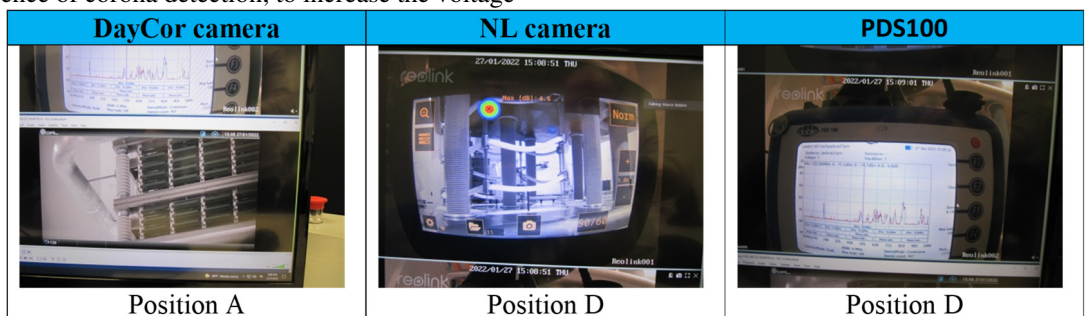

Fig. 7 Test at 500 kV.

After TPS repair and general plant inspection, the second test session took place from March 28$^{th}$ until April 1$^{st}$ 2022. The diagnostic system was improved with:

− two additional capacitive field probes, for a total of 4: 3 of them located on the floor in correspondence of three HVD1 sides ("C1", "C2", "C4" in Fig. 2), the fourth ("C3") below the DL; the probes, which signals were sent through coaxial cables to the four inputs of an oscilloscope located very close to probe "C2", are represented in Fig. 2 with the same colours used for the oscilloscope traces;

− a camera in the visible field looking at the SCD inside the vessel and other electrical measurements (voltage dividers and current transducers) distributed along the plant;

− the display of TPS voltage and current signals on a screen in the control room, together with the trend of the electrical conductivity of the demineralized cooling water, flowing inside the HVD1 insulated breaks [4] (too high conductivity makes the TPS to loose voltage control).

It was agreed to deviate from the test procedure: the voltage was raised in step of 100 kV up to 500 kV by keeping the PD diagnostic instrumentation in only one position ("D" in Fig. 2); for all the other positions tests started directly from a voltage level of 600 kV.

The analysis of the field probe signals allowed to identify <u>three typical patterns of the records</u>, though not regularly reproducible in each position and for exactly the same voltage level:

- <u>oscillation of the four field probe signals</u> (see Fig. 8): the four signals oscillated with a "period" of about 1.8 s but the signal of probe 3, closer to DL, had smaller amplitude; the oscillations could be correlated with the regulation of the voltage applied to HVD1 by the TPS, measured by the probe acting as low voltage branch of a voltage divider completed by its stray capacitance towards the HVD1;

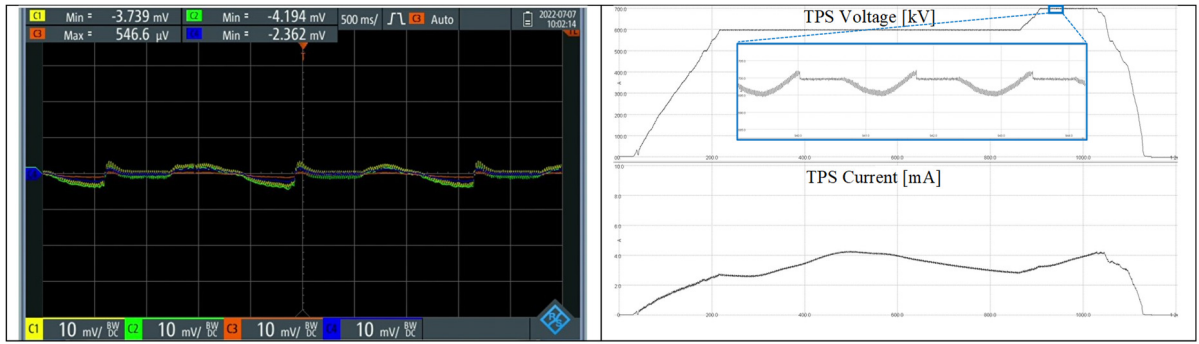

Fig. 8 Field probes (oscillation at 700 kV) and TPS V-I signals.

- <u>field probes signals identical with almost null value</u>: after investigation, it was clarified that this behaviour occurred when, due to the water conductivity increase because of its degradation, TPS current exceeded the set threshold and consequently output voltage slightly decreased (see red rectangle in Fig. 9, switch of TPS from voltage to current control);

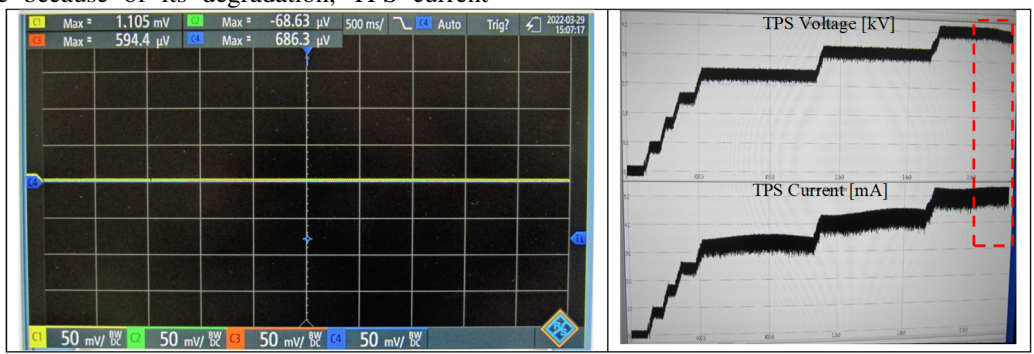

Fig. 9 Field probes (600 kV after 10 min) and TPS V-I signals.

- <u>Shift of one field probe signal</u>: starting from about 700 kV a shift of the signal of field probe C2 was randomly noticed, while the other field probes signals continued to oscillate around zero (see Fig. 10 left). The amplitude of the shift of the C2 trace was not constant, ranging from 1 to 15 mV without any apparent correlation with TPS applied voltage. Some hypothesis was done to explain the shift of the capacitive probe signal (the nearest one to the HVD1 insulating pipes for cooling water): corona effect, continuous variation of the electric field distribution due to no-ionizing phenomena, measuring error. By analysing the measurements of the other PD diagnostics (NL camera and PDS100) in different observation positions no evidence of corona was found (see Fig. 10 centre and right for position C), therefore corona effect indication by the probe was deemed unlikely and the understanding of the reason of the shift was left to further verifications.

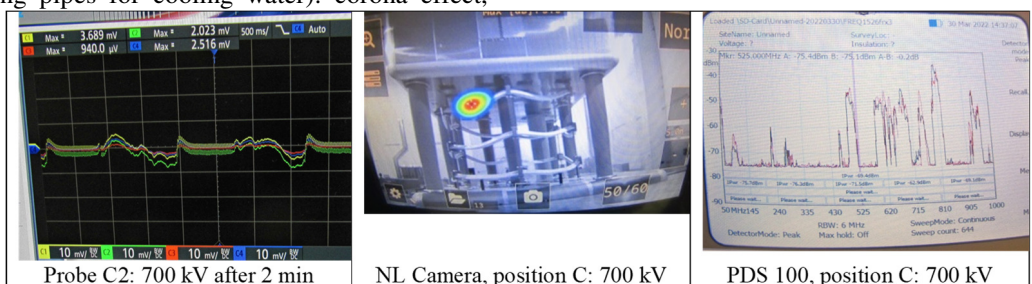

| Probe C2: 700 kV after 2 min | NL Camera, position C: 700 kV | PDS 100, position C: 700 kV |

Fig. 10 C2 field probe (shift at 700 kV) and PD measurements.

During this session a voltage of 900 kV was achieved. The aim was to maintain the voltage for 10 min but the test was stopped after few minutes because of too high current threshold reached by TPS. After restoring low water conductivity, during the new rise phase a great noise from the HVH was heard at 870 kV and the voltage dropped to zero. Moreover, the camera looking at SCD captured a flashover inside the vessel. The diagnostic systems did not show any indication of pre-discharge before failure (Fig. 11).

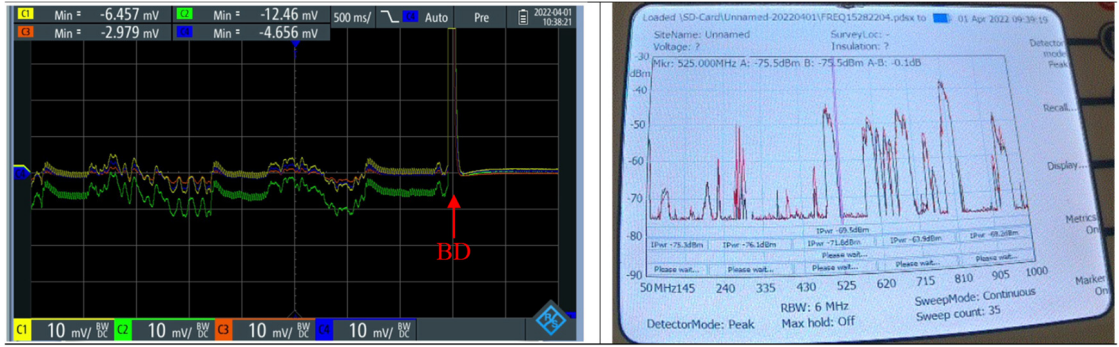

Fig. 11 Field probes and PDS 100 (position D) at BD instant.

Only NL Camera (see Fig. 12, temporal sequence from left to right) allowed to identify during post analysis the change of the source of sound: before the BD the usual sound sources are visible, while at the flashover a new important source of sound was indicated at the penetration of the HVBA in the HVD1 (sound intensity is reported on display top position). The comparison between the frames sequence indicates that the sound propagated from the HVBA to the HVH and not vice-versa.

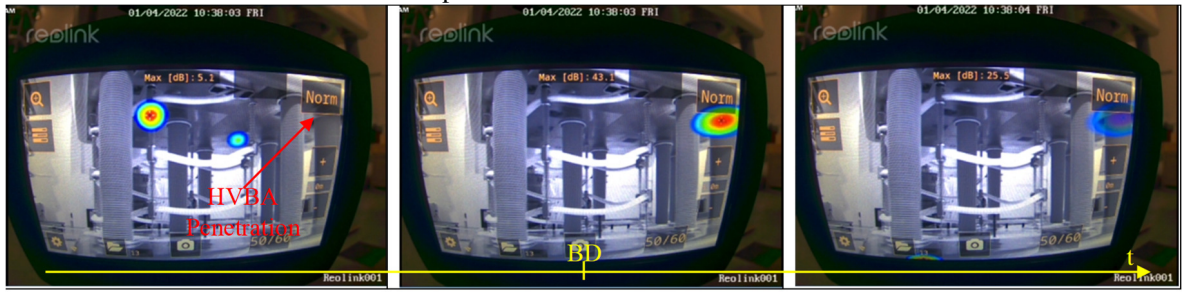

Fig. 12 NL Camera (posit. D) frames sequence at BD event.

The information gathered were conflicting and didn't allow for a clear interpretation of the event, in fact:

- no detection of pre-discharge phenomena by PDS100 and NL camera;
- no ionizing phenomena revealed by DayCor camera, that together with the previous item observations indicate as unlikely the possibility of a BD in air insulated equipment;
- images acquired by the camera looking inside the vessel confirmed that an arc occurred between the electrodes of the SCD;
- the arc between the SCD electrodes was also confirmed by a current measurement in series with the SCD ground connection.

  Nonetheless:

- the arc on the SCD may have been induced by an overvoltage due to a main BD located somewhere else in the plant (secondary BD);
- the indication of strong sound propagating from the HVBA penetration inside HVD1 towards the HVH may suggest that the BD occurred in the HVH surroundings, not necessarily in air;
- the information that some activities was detected in TL2 and TL3 before BD by the embedded PD detectors (system not under direct online monitoring) enforced the hypothesis of primary breakdown in $SF_6$ insulation.

Anyway, the inspection carried out in air insulated equipment as well as inside the gas insulated components did not reveal signs of discharge.

To improve the phenomena understanding (dedicated circuit modelling and simulations results are described in [14] [15] [16]), a third test campaign has been carried out from 12$^{th}$ to 18$^{th}$ July 2022. SCD fixed spark gap distance was increased, being the previous configuration very close to the nominal (-1 MV) voltage limit (the fixed spark gaps were indeed not present during voltage withstanding test 5, see Sec. 2). The diagnostic system was further improved adding:

− one camera in the visible field provided with microphone ("7+M" in Fig. 2 and Fig. 3) to detect any occurrence of light or sound produced inside the HVD1;
− a microphone installed nearby the vessel ("M" in Fig. 3) to detect the sound occurrence generated by a discharge inside the vessel;
− high speed camera looking at the SCD fixed spark gap,
− 6 ambient microphones installed along the plant trying to identify the location of the sound occurrence associated to a BD event and its propagation along the plant,
- other electric measurements (mainly current transducers around TL tanks).

To understand the reason of the shift of the signal of field probe C2 in the previous test, the session started with all the PD diagnostic instruments (PDS100, NL and DayCor cameras) installed in position C, in front of the probe, to discriminate the signal drift due to corona inception from signal drift due to coupling or interference. The test voltage was gradually increased in steps of 100 kV, maintaining the application for some minutes every time, up to 700 kV lasting for 10 min. No shift at all was detected on the signal probe, which showed only the voltage ripple applied to HVD1 by the TPS (see Fig. 8 for reference).

Such voltage level was confirmed also by moving the instrumentations in position D and B, ramping up directly to 600 kV and then 700 kV, applying both voltages for 10 min. By keeping the instrumentations in position B, it was achieved 900 kV applied for 23 min, the longest time possible before TPS high current threshold intervention. During the voltage application, some PD temporary activity was detected by PDS100 at around 500 MHz and some dots were spotted by DayCor camera and the signals recovered to background noise level after some minutes.

The test has been repeated with lower water conductivity; the voltage was increased linearly (~100 kV / min) to 500 kV, then applied for at least 2 min and the same procedure was repeated at 600 kV and 700 kV; then in step of 50 kV, kept for 2 min, the voltage was raised up to 850 kV but after 2 s collapsed because of a breakdown.

As for the previous test session, data analysis didn't allow for unambiguous interpretation, because:

- no precursors of the BD on air insulated equipment were detected by PDS100, NL and DayCor cameras;
- images acquired by the still cameras looking inside the vessel as well as the current measurement in series with the SCD ground connection confirmed that an arc occurred and lasted some hundreds of microseconds between the SCD electrodes; any corona occurrence associated to this current, had to be detected by the dedicated diagnostics;
- a strong sound was registered inside the HVH (but the new camera "7+M" allowed to exclude that it was originated inside the HVD1) as well as nearby the vessel by the new microphone "M";
- the sound was captured by the 6 ambient microphones but at the same starting time and all the microphones saturated, making therefore impossible to discriminate the source position;
- some PD activities was detected mainly in TL2 before BD by the embedded detectors;
- additional electric measurements patterns along the plant are compatible with a discharge between the high voltage conductor(s) and the return line, a discharge likely occurring in the $SF_6$ insulated equipment.

All these findings didn't allow to understand whether the SCD arc was the main event or if it was consequent to a BD happened elsewhere in the plant. Nonetheless, based on PD diagnostics measurements and HVH cameras registration, it can be excluded that the BD occurred on air insulation.

Visual inspection carried out in the gas insulated components detected discharge marks on SCD fixed spark gaps, mainly on 1 – 0.8 MV and 0.8 – 0.6 MV stages; no other signs of discharge have been revealed in the remaining $SF_6$ insulated components as well as in the air insulated one.

## 5 Conclusions

MITICA integrated commissioning started in 2021 allowing to reach smoothly 700 kV x 1000 s at a first attempt in such large NBI, an important result in view of ITER/DEMO; nevertheless, two breakdowns occurred somewhere in the HV plant, either in air or in SF6, without clear evidence of the source provided by the existing set of diagnostics.

To identify the locations of possible weak insulation points responsible of the voltage breakdown occurrence, the PD detection in MITICA has been improved on air-insulated parts by identifying a set of suitable instrumentation amid off-the-shelf solutions in AC applications. Dedicated activities to verify the correct operation of the instruments for PD detection in DC and their field of view were carried out.

Considering the huge space to monitor inside the HVH and the effective field of view of the instruments, the diagnostic equipment was installed to monitor a limited area; in total 4 points were selected to monitor the complete installation inside the HVH. Four capacitive field probes were disposed on the floor, in correspondence of three HVD1 sides whilst the fourth below the DL.

Three test campaigns were carried out in 2022. The information gathered during the tests resulted not conclusive but, based on PD diagnostics measurements, at least it has been concluded that very unlikely the BD occurred in the air insulated equipment.

Further test campaigns are scheduled in MITICA in 2023. Being the identified PD instrumentation proven to be effective in DC air diagnostic, it will be studied and optimized a definitive installation (in MITICA as well as in future in ITER HNBs) for detection and protection against system BD. Moreover, the possibility to implement a non-invasive PD detection on $SF_6$ insulated components (by means of sensors to be applied on the equipment tanks) will be investigated in preparation of the incoming tests campaign, to improve weak insulation points detection necessary either for the test campaign or to gain system reliability during the future HNBs operation.

MITICA is demonstrating its mission to identify and address unpredictable problems in the integration of high voltage non-standard components. The collected lessons learnt in the handling of the unprecedented high voltage levels will facilitate the future operation in ITER.

**Acknowledgement and Disclaimers**

This work has been carried out within the framework of the ITER-RFX Neutral Beam Testing Facility (NBTF) Agreement and has received funding from the ITER Organization. The views and opinions expressed herein do not necessarily reflect those of the ITER Organization. This work has been carried out within the framework of the EUROfusion Consortium, funded by the European Union via the Euratom Research and Training Programme (Grant Agreement No 101052200 — EUROfusion). Views and opinions expressed are however those of the author(s) only and do not necessarily reflect those of the European Union or the European Commission. Neither the European Union nor the European Commission can be held responsible for them.